# What is 3C 324?

*Mark Dickinson[1], Arjun Dey[2,3] and Hyron Spinrad[2]*

[1] Space Telescope Science Institute, 3700 San Martin Dr., Baltimore MD 21218, USA
[2] Department of Astronomy, Univ. of California, Berkeley CA 94720, USA
[3] Institute of Geophysics and Planetary Physics, LLNL, Livermore CA 94550, USA

## 1 Radio Galaxies at High Redshift – Hopes and Fears

At one time, radio sources offered our only access to the universe of galaxies at very large redshifts. Amongst the swarms of faint objects in the sky, powerful radio emission draws our attention and guides our telescopes. Strong emission line spectra allow us to measure redshifts out to $z = 4.25$ (thus far – Lacy *et al.* 1994). The majority of extragalactic, steep spectrum radio sources have faint, spatially extended optical counterparts, and exhibit narrow line emission. Thus we identify them as galaxies, regardless of what AGN activity may have drawn our attention in the first place.

Nearby, powerful radio galaxies are giant ellipticals. This association led to the hope that their high–redshift counterparts could be studied as progenitors – that their evolution could be traced through their colors and magnitudes. The small photometric scatter observed in the infrared Hubble diagram (magnitude vs. redshift relation) out to at least $z \approx 2$ has suggested that radio galaxies are a stable population of intrinsically similar galaxies, perhaps even well–enough behaved for use as cosmological probes (Lilly 1989, McCarthy 1993).

At the same time, some properties of high–$z$ radio galaxies appear to be quite extraordinary. Their optical (i.e. rest–frame UV) magnitudes and colors vary widely (cf. Lilly and Longair 1984), and high–quality images often reveal highly elongated morphologies, sometimes resolved into multiple clumps (cf. Le Fèvre and Hammer 1988).

A breakthrough (or death–blow, for the pessimist) occurred when it was realized that the elongated, complex morphologies closely share the primary axis of the radio source (McCarthy *et al.* 1987, Chambers *et al.* 1987). This was clear evidence that the radio source, a supposedly short–lived phenomenon of the central AGN, exerts a dramatic influence on the host galaxy. Early interpretation (see McCarthy 1993 for references and a review) of this 'alignment effect' centered on massive star–formation induced by the passage of the radio jet. It was suggested that perhaps a large fraction of the stellar mass could be formed in this fashion: perhaps these are true protogalaxies, caught in the act of formation.

However, the discovery of strong linear polarization in many objects cast doubt about whether the light we see is starlight at all (di Serego Alighieri et al. 1989).

Since the most intriguing and unexplained aspect of high–$z$ radio galaxies is their aligned morphologies, it has long been hoped that imaging with the *Hubble Space Telescope (HST)* would help explain the origins of these peculiar structures by elucidating their kiloparsec–scale details (cf. Miley et al. 1992). Longair et al. (1995) and Röttgering (this volume) have presented initial results from an *HST* imaging survey of high redshift 3CR galaxies. Here we will examine in detail the properties of a single, prototypical example of the alignment effect, taking advantage of a uniquely deep *HST* image, as well as infrared imaging and deep optical spectroscopy from the Keck 10m telescope.

## 2 The Case of 3C 324

Figure 1 shows deep optical ($R$) and infrared ($K$) images of 3C 324, an aligned radio galaxy at $z = 1.206$. In the rest–frame, the $R$–band image samples the galaxy light in near–ultraviolet, while the $K$–band provides an image at $\sim 1\mu$m. In the infrared, 3C 324 appears to be a round, giant elliptical galaxy. In the UV, it is bimodal (at least) and elongated in the general direction of the radio source. Optical polarimetric observations (di Serego Alighieri et al. 1993, 1994; Cimatti et al. 1995) measure a polarization of 11–18% for 3C 324, with the electric vectors perpendicular to the major axis of the aligned light.

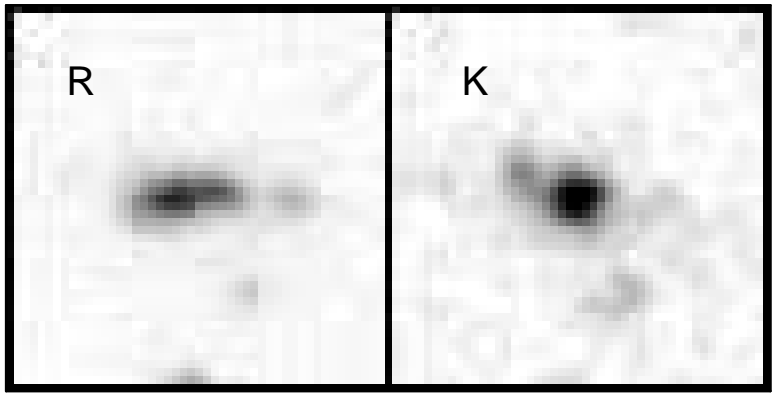

**Fig. 1.** $R$ and $K$–band images of 3C 324 ($z = 1.206$), obtained with the Keck and KPNO 4m telescopes, respectively. The field of view is $10'' \times 10''$ for each panel, and the seeing FWHM=$0\rlap{.}''85$ for both.

Figure 2a presents an *HST* image of 3C 324 with 5 GHz radio contours superimposed. The galaxy was observed with the Planetary Camera (PC) and F702W filter ($\lambda_{\text{eff}} \approx 7000$Å). A very long exposure (32 orbits, $t_{\text{exp}} = 18$ hours) was obtained in order to study extremely faint galaxies in a cluster surrounding the radio source. The PC pixel scale, $0''\!.046$/pixel, corresponds to $\sim$500 pc at the redshift of the radio galaxy (for $H_0$=50 km s$^{-1}$ Mpc$^{-1}$, $q_0$=0.1). The F702W filter samples the galaxy light at $\sim$3200Å in its rest-frame. Although several emission lines are included in this bandpass (MgII, [NeV]), the strongest lines (e.g. [OII]3727) are largely excluded. Our spectra (see below) show that line emission contributes only 11% to the integrated flux from the galaxy, although the point-by-point contribution could be smaller or larger depending on the spatial distribution of the emission line gas.

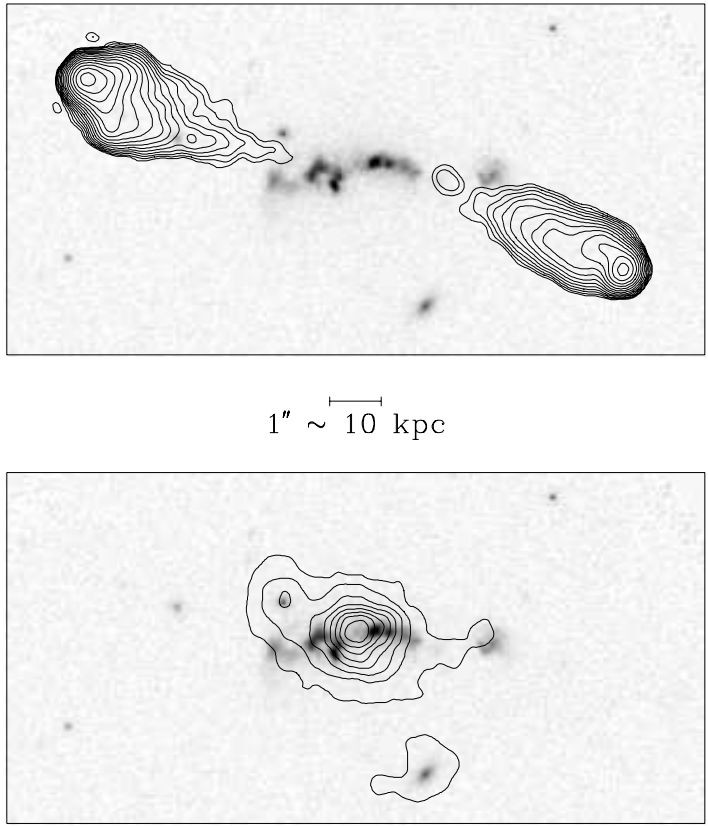

**Fig. 2.** *(a) Top: HST* Planetary Camera image of 3C 324 through the F702W filter. The contours show the 5 GHz VLA map of Fernini *et al.* 1993. The formal uncertainty in radio-optical registration is $\sim$$0''\!.15$ in each coordinate, although systematic errors in the astrometry could be larger. *(b) Bottom:* KPNO 4m $K$-band image superimposed on the *HST* image. The peak of the $K$-band light lies between the major complexes of rest-frame UV emission.

At ∼ 0.″1 resolution, the elongated structure of 3C 324 breaks up into a long, remarkably narrow chain of compact, high–surface brightness clumps, with scale sizes ∼0.″2 to 0.″4 (FWHM), i.e. ∼2 to 4 kpc. There is little point–by–point correspondence between radio and optical features. Considering the axes defined by the radio hotspots, the radio jet (visible on the eastern side), and the UV continuum, there is a sense of counterclockwise rotation as one progresses from large angular scales to small, suggesting a precession of the AGN emission axis.

Figure 2$b$ overlays the PC image with contours from the $K$–band image. While the angular resolutions of the *HST* and ground–based infrared data differ substantially, it is evident that the $K$–band light peaks at a local *minimum* in the rest–frame UV emission, between the two major "blob complexes." If this is considered to be the center of the host galaxy, it is apparently heavily obscured at ultraviolet wavelengths.

Figure 3 plots a spectrum of 3C 324 taken with the Keck Telescope and LRIS spectrograph. The slit was oriented along the optical major axis, and the extraction presented here sums nearly all of the light from the galaxy. Several features are notable:

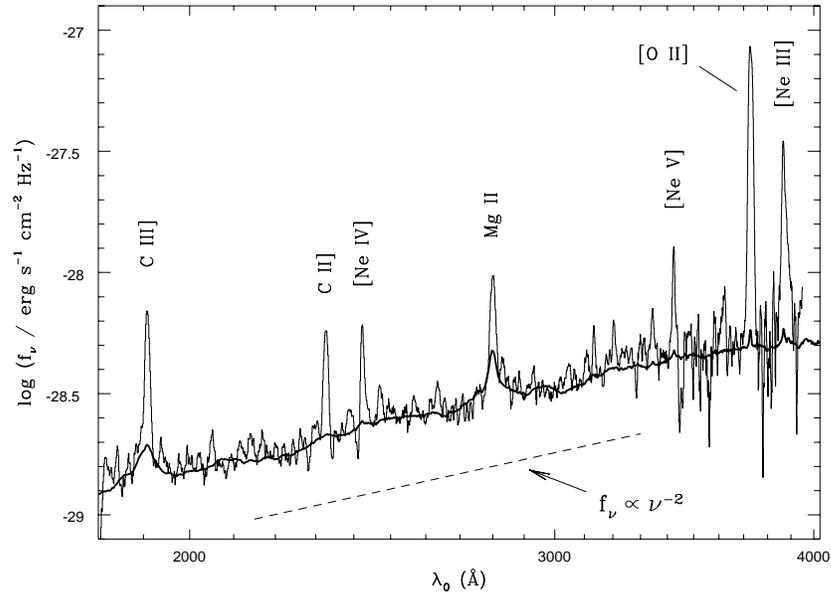

**Fig. 3.** Log–log plot of a total light spectrogram of 3C 324, obtained with the Keck LRIS. The heavy line superimposes the average LBQS QSO spectrum of Francis *et al.* 1991, artificially reddened to approximately match the slope of the radio galaxy continuum. The MgII and CIII] emission lines in the radio galaxy exhibit broad wings consistent with the presence of underlying, quasar–like broad lines with normal QSO equivalent widths.

- The aligned galaxy continuum is actually quite red. It is reasonably well represented by a power law $f_\nu \propto \nu^\alpha$ with spectral index $\alpha \approx -2$.
- The permitted and semi–forbidden lines (MgII, CIII]) which dominate quasar spectra appear to have broad wings. In figure 3 we superimpose a composite QSO spectrum over that of the radio galaxy. In order to match the red continuum slope of the radio galaxy, we have artificially "reddened" the quasar by 1.6 powers of $\nu$ and scaled the continuum to that of the radio galaxy at $\sim 3000$Å. The MgII and CIII] lines of the quasar provide a good match to the broad wings seen in the radio galaxy spectrum. The broad features cannot arise from an active nucleus viewed directly, since no such unresolved feature is observed in the *HST* images. Indeed the presumed location of the nucleus appears to be heavily obscured in figure 2*b*. Note that strong, narrow MgII and CIII] emission is superimposed on the broad features, which would partially mask their visibility in lower signal–to–noise data.

## 3 Interpretation

The strength and position angle of the observed polarization in 3C 324 strongly suggests that some fraction of the UV continuum is scattered. Other mechanisms for producing the polarized light seem untenable. E.g. for synchrotron radiation, one would expect better correspondence between radio and optical components unless the two emission regions produce radically different synchrotron spectra. Moreover, the broad wings seen in MgII and CIII] resemble quasar emission lines with normal equivalent widths, supporting the assumption that most or all of the radio galaxy continuum is scattered AGN light. Previous objections to scattering models based on the lack of broad line emission need not apply to 3C 324 at least. Substantial contribution from additional *in situ* continuum would only dilute the equivalent widths of the scattered broad components, requiring that their *intrinsic* strength be greater than that seen in typical quasars. Similarly strong, broad MgII features have been observed in other radio galaxies, and evidence has been presented demonstrating that MgII may be polarized (e.g. di Serego Alighieri *et al.* 1994) and spatially extended (Dey and Spinrad 1995) – both clear predictions of the scattering hypothesis.

The surface brightnesses of the clumps in the *HST* image decline with increasing distance from the galaxy center and presumed AGN location – see figure 4. This is qualitatively consistent with expectations from a scattering model, although the prediction is not unique. The spatially resolved [OII] emission in our long–slit spectra is strongest at the linear extremes where the continuum surface brightness is lowest, contrary to what might be expected if the continuum morphology and line emission traced recent star formation along the jet axis. In addition, this provides further evidence that line emission is not dominating the morphology of the *HST* image.

Where is the giant elliptical host galaxy? It apparently dominates the *K*–band image, but is seemingly invisible at $\lambda_{\rm obs} = 7000$Å in the *HST* data. To quantify this, we have simulated a gE host galaxy at $z = 1.2$. An actual brightest cluster elliptical (BCE) at $z = 0.4$ observed with *HST* by Dressler *et al.* (1994) was artificially "redshifted" out to $z = 1.2$. Angular size was converted assuming $q_0 = 0.5$, and surface brightnesses were set by requiring the host galaxy to have $K = 17.5$ (as we measure from the $K$ images) and $(R - K) = 5.9$, the color of other faint ellipticals in the cluster surrounding 3C 324 (cf. Dickinson 1995 in this volume). Noise was added to match the real PC data. The result is shown in figure 5. The low surface brightness of the BCE envelope makes it difficult to detect, but the galaxy nucleus should be evident. No such feature is seen in the actual image (cf. figure 2*b*), however. We conclude that either the host galaxy is substantially less nucleated than a typical BCE, or that it is obscured by dust. We calculate that extinction with $E(B - V) \approx 0.3$ is sufficient to render the galaxy nucleus invisible in the PC image. This also provides further evidence for the presence of distributed dust in 3C 324, which can also serve as the scattering medium for producing the aligned UV light.

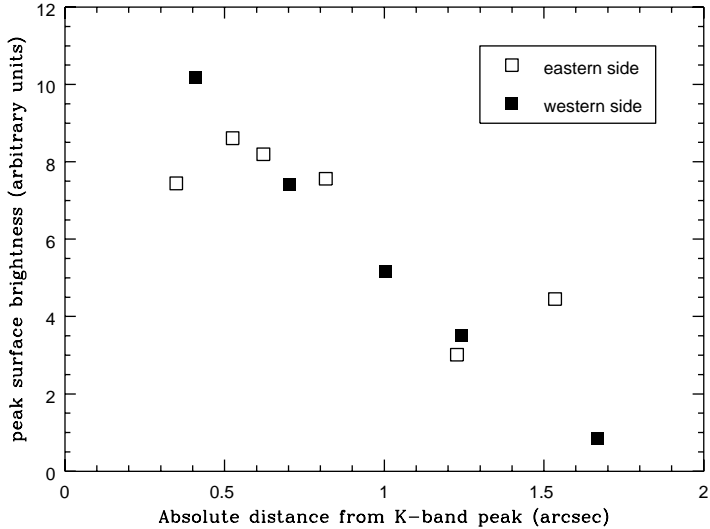

**Fig. 4.** Peak surface brightnesses (averaged over $0''\!.3$ diameter apertures) of the clumps in 3C 324 vs. their radial distances from the $K$–band peak. A nearly monotonic decline is observed, particularly in the tight, curving chain of knots to the west of the galaxy center.

There seems to be no hard evidence against a pure–scattering model for the aligned UV continuum in 3C 324. The present observations cannot confirm or firmly exclude other possible sources, e.g. hot stars, hydrogen recombination continuum, etc. These should be unpolarized, however, and high quality polarimetry (imaging and spectroscopic) should go a long way toward providing a resolution. Keck spectropolarimetry is now in hand (Cimatti *et al.* 1995 *in prep.*), and WFPC2 imaging polarimetry with *HST* is scheduled for Cycle 5.

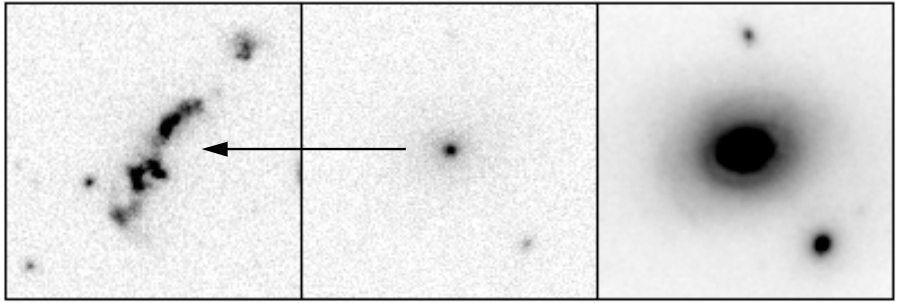

**Fig. 5.** Simulation of a giant elliptical host galaxy for 3C 324. The left panel shows the *HST* Planetary Camera image of the radio galaxy. An *HST* observation of a brightest cluster elliptical (BCE) at $z = 0.4$ is seen at right. The center panel simulates the BCE as it would appear at $z = 1.206$ in the Planetary Camera field.

Several tasks remain before a model for the alignment effect in 3C 324 can be considered convincing:

- It must explain the continuum configuration seen in the *HST* images. Why are the blobs so compact and colinear? AGN unification models such as that of Barthel (1989) propose a anisotropic "radiation cone" with an opening angle of $\sim 45°$. Let us suppose that the AGN lies at the $K$-band peak, and shines out in a cone roughly axisymmetric about the radio jet axis. The UV light we observe is certainly confined within such a cone, but does not fill it. Rather, it forms a linear configuration, clumped along the long axis. At first glance, this arrangement may seem to fit more naturally with a cartoon picture of jet-induced star formation than with scattering models. If the bulk of the UV light is scattered, then apparently the scattering medium is also strongly clumped – the "mirrors" are closely confined to an axis which is similar (but not quite identical to) the radio source axis. Does the radio jet play a role in creating this clumpiness? If the UV light is emitted *in situ* by sources along this axis, we are then left to explain the observed polarization and broad line emission.
- Why is the aligned light red? If it is scattered from a hidden but otherwise ordinary central quasar, it must have been reddened. Extinction at the source (e.g. as the AGN shines out through a dusty medium closely surrounding the galaxy center) is possible, but would strongly attenuate ionizing UV photons, posing a difficulty for AGN photoionization models of the extended emission line gas. While simple, optically thin Rayleigh scattering would make the incident spectrum *bluer*, realistic models produce more complex effects, due to multiple scattering, absorption, and varying albedo depending on the grain-size distribution (e.g. Laor 1995 *in prep.*) Similar problems face other models for the continuum emission. Spectrophotometry and high angular resolution color measurements over a wide range of wavelengths would be invaluable for constraining the nature of the continuum.
- If the light is scattered, what is the scattering medium? Dust and electrons are the two popular contenders. The apparent obscuration of host galaxy nucleus and the red color of the aligned UV continuum suggest that dust may be abundant. If dust is responsible for scattering, what is its origin, and why is it distributed throughout a volume tens of kpc across? Is it manufactured in the host galaxy and strewn wide by outflows/winds? Could it precipitate from the surrounding cluster, either from cooling intra-cluster gas or from companion galaxies? Or is the dust somehow manufactured by the radio source itself?

For those who hope to study radio galaxies as *galaxies*, these conclusions may perhaps be disappointing. It appears that the optical light from objects like 3C 324 is entirely dominated by a manifestation (albeit indirect) of an AGN phenomenon. While the alignment mechanism still presents us with fascinating puzzles to solve, its relevance for understanding galaxies, and correspondingly the utility of radio galaxies themselves for studying evolution, is questionable.

The optimist, however, might consider these new results encouraging. If the aligned light is only a reflection of the AGN, then it is in some sense a phantom. Perhaps the underlying galaxy is largely unaffected by the "fireworks," contrary to extreme star-formation scenarios for the alignment effect where the stellar content of the galaxy could be dramatically altered as a consequence of the AGN activity. The infrared properties of radio galaxies, where the aligned light is apparently least important, remain reasonably unaffected and are still a valid subject for evolutionary studies, e.g. via the infrared Hubble diagram.

Moreover, recent work suggests that the optical alignments may be weaker or absent in galaxies with less powerful radio sources (Dunlop and Peacock 1993, but cf. McLeod 1994 for a contrary view). Perhaps this phantom only haunts those few, rare galaxies at the extreme bright end of the radio luminosity function – even the optical properties of weaker radio sources might be safe for study.

## 4 Acknowledgements

We would like to thank the conference organizers for providing a wonderful opportunity and locale for this gathering, and the castle staff for keeping the Hexenzimmer well stocked with wine. We also thank the other co-investigators on our *HST* imaging program: George Djorgovski, Peter Eisenhardt, and Adam Stanford. M.D. acknowledges generous travel support from STScI.